\documentclass[useAMS,usenatbib]{mn2e}

\usepackage{graphicx}

\newcommand{\civ}{\hbox{C\,{\sc iv}}}  \newcommand{\nv}{\hbox{N\,{\sc
      v}}}  \newcommand{\mgii}{\hbox{Mg\,{\sc ii}}}
\newcommand{\oiii}{\hbox{O\,{\sc iii}}} \newcommand{\hi}{\hbox{H\,{\sc
      i}}} \newcommand{\kms}{\hbox{km\,s$^{-1}$}}
\newcommand{\W}{$W_0^{\lambda1548}$}

\title[Velocity
  Distribution of Narrow C\,{\normalsize \it IV} Absorbers]
  {The Quasar-frame Velocity Distribution of Narrow
   C\,{\Large\bf IV} Absorbers} 
\author[Nestor, Hamann \& Rodriguez Hidalgo]
       {D.~Nestor,$^{1,2,\star}$ F.~Hamann,$^2$ and
         P.~Rodriguez Hidalgo$^2$ \\ $^{\star}$ dbn@ast.cam.ac.uk
         \\ $^{1}$ Institute of Astronomy, Madingley Road, Cambridge
         CB3 0HA, UK\\ $^{2}$ Department of Astronomy, University of
         Florida, Gainesville, FL, 32611, USA\\ } \date{Accepted
         Received  }

\begin{document}

\pagerange{\pageref{firstpage}--\pageref{lastpage}} \pubyear{}

\maketitle

\label{firstpage}

\begin{abstract}  
We report on a survey for narrow (FWHM $< 600$ \kms) \civ\ absorption
lines in a sample of bright quasars at redshifts $1.8 \le z < 2.25$ in
the Sloan Digital Sky Survey.  Our main goal is to understand the
relationship of narrow \civ\ absorbers to quasar outflows and, more
generally, to quasar environments.  We determine velocity zero-points
using the broad \mgii\ emission line, and then measure the absorbers'
quasar-frame velocity distribution.  We examine the distribution of
lines arising in quasar outflows by  subtracting model fits to the
contributions from cosmologically  intervening absorbers and
absorption due to the quasar host galaxy or  cluster environment.  We
find a substantial number ($\ge 43\pm6$ per cent) of absorbers with
\W\ $> 0.3$ \AA\ in the velocity range +750 \kms\ $\la v \la $ +12000
\kms\ are intrinsic to the AGN outflow.  This `outflow fraction' peaks
near $v=+2000$ \kms\ with a value of $f_{outflow} \simeq 0.81 \pm 0.13$.  At
velocities below $v \approx +2000$ \kms\ the incidence of outflowing systems
drops, possibly due to geometric effects or to the over-ionization of
gas that is nearer the accretion disk.  Furthermore, we find that
outflow-absorbers are on average broader and stronger than
cosmologically-intervening systems.  Finally, we find that $\sim 14$
per cent of the quasars in our sample exhibit narrow, outflowing \civ\
absorption with \W\ $> 0.3$\AA, slightly larger than that for broad
absorption line systems.  
\end{abstract}


\section{Introduction}
\label{Sec:intro}
High speed outflows observed in rest-frame UV absorption lines are
common  components of active galactic nuclei (AGN). The ejected gas is
believed to be driven off of the accretion  disk around a central
super-massive black hole (SMBH) by either  radiation pressure (Proga
et al.\ 2000 and 2004) or  magneto-centrifugal `forces' (Everett
2005). In luminous quasars, the outflows might be sufficiently
energetic to provide important feedback to the galactic surroundings,
potentially disrupting star formation and halting further growth of
the SMBH (Hamann \& Ferland, in prep., Hamann et al.\ 2007, Di Matteo
et al.\ 2005, and refs. therein).  Quasar outflows might also
contribute to the blowout of gas and dust from young galaxies, and
thereby help to distribute metals to the surroundings and reveal the
central accreting SMBH as an optically  visible quasar (Silk \& Rees
1998, Moll et al.\ 2007). 

Nonetheless, AGN outflows remain poorly understood. Most quasar
studies have focused on the blueshifted broad absorption lines (BALs),
which very clearly identify powerful outflows at typical  velocities
of 5000 to 20000 \kms\ (Korista et al.\ 1993, Weymann et al 1991).
BALs are detected in $\sim$10 per cent of optically-selected quasars
(Trump et al.\ 2006), but the flows themselves might be ubiquitous if,
as expected, the absorbing gas subtends just a fraction of the sky  as
seen from the central continuum source (e.g., Hamann et al.\ 1993). 

There is, however, a wider range of observed outflow phenomena than
just the BALs.  For example, there are the so-called `associated'
absorption lines  (AALs) that appear in roughly 25 per cent of quasar
spectra (see  also Vestergaard 2003). These features are much narrower
than the BALs, with full widths at half minimum (FWHMs) of several
hundred \kms\ or less, and they have  redshifts that are, by
definition, within a few thousand  \kms\ of the emission line
redshift, i.e., $z_{abs}\approx z_{quasar}$  (e.g., Foltz et al.\
1986, Anderson et al.\ 1987). However,  unlike the BALs, the
identification of AALs with quasar outflows  is often
ambiguous. Weymann et al.\ (1979) noted that AALs can have a  variety
of origins, including quasar host galaxies or cluster  environments,
and cosmologically intervening (unrelated) material,  in addition to
quasar outflows. Nonetheless, there is growing  evidence that a
substantial fraction of AALs and other narrow absorption lines (NALs)
at even larger velocity shifts do  form in quasar outflows. In their
sample of 66 quasar spectra, Weymann et al.\ measured a strong excess
of \civ\ systems with $z_{abs} \simeq z_{quasar}$ compared to $z_{abs}
<< z_{quasar}$.  Although the statistics were somewhat poor, they
measured a quasar rest-frame velocity width for this excess of
$\approx 1200$ km s$^{-1}$, centred near $v=0$, which they attributed
to the quasar cluster environment, and a significant tail extending
out to blueshifts of $\approx 15000$ km s$^{-1}$ which they identified
with an ejected/outflow component. 

More recently, Richards (2001) found a $\sim$36 per cent excess of
narrow \civ\ absorbers with $v > 5000$ \kms\ in flat-spectrum quasars
as compared to steep-spectrum quasars.  This was interpreted as
evidence that over a third of high-velocity NALs are ejected from
quasars. Well-studied individual cases provide direct evidence for an
outflow origin of some NALs based on, for example, variable  line
strengths, resolved line profiles that are smooth and very broad
compared to the thermal speeds, and doublet strength ratios showing
that the absorber covers only part of the background light source
along  our observed sightlines (e.g., Hamann et al.\ 1997a and 1997b,
Barlow \& Sargent  1997, Misawa et al.\ 2007, and refs.  therein). Two
variability studies, in particular,  indicate that at least $\sim$21
per cent of quasar AALs form in the inner, dynamic  regions of quasar
outflows (Narayanan et al.\ 2004, Wise et al.\ 2004). 

Determinations of the detection frequency and the velocity
distribution of ejected NALs are essential to our understanding  of
the basic structure and physics of quasar outflows.  These results
will also impact studies of intervening systems.  Cosmologically
intervening \civ\ absorbers are used to study galaxy halos and the
enrichment history of the IGM (e.g., Songaila, 2001;  Ryan-Weber,
Pettini \& Madau, 2006; etc.)  Determining the contamination of 
intervening samples by intrinsic absorbers is therefore pertinent to
those studies.

With the goal of mapping the incidence of narrow \civ\ absorption in
quasar rest-frame velocity space, we make use of the large publicly
available spectroscopic database from the Sloan Digital Sky Survey
(SDSS; York et al., 2000).  In \S\ref{Sec:data} we describe our data
set.  In \S\ref{Sec:analysis} we describe the analyses of the absorber
catalog and possible systematic effects, in \S\ref{Sec:disc} we
discuss the implications of our results and in \S\ref{Sec:conc} we
summarize our conclusions.

\section{Data Set}
\label{Sec:data}
\subsection{The Quasar Sample} 
We selected quasar spectra from the 4th data release of the SDSS.  The
spectra cover a wavelength range of 3820\AA\ (with resolution  R
$=1800$) to 9200\AA\ (R $=2100$), corresponding to \civ\ redshifts of
$z_{abs} > 1.468$ and resolutions of 143 \kms\ $\le \delta v \le 167$
km s$^{-1}$.  We wish to study absorbers over a large span of
quasar-frame velocities, from negative velocities (i.e., redshifted
with respect to the quasar) to large ($\approx 71000$ \kms) positive
velocity corresponding to the red-edge of the  Ly$\alpha$
emission.\footnote{Absorber-quasar velocity differences greater than 
$\simeq 71100$ \kms\ place the \civ\ $\lambda1548$ absorption  
feature in the Ly$\alpha$ forest, making
identification and measurement prohibitively
difficult.}  To cover this complete range requires $z_{quasar}>2.14$.
However, in order to accurately measure the quasar redshifts (see
\S\ref{Sec:mgz} below), we require $z_{quasar} < 2.25$ so that the
quasar \mgii\ emission line is covered by the data.  Furthermore, the
number of high signal to noise ratio (S/N) spectra available decreases
with increasing redshift.  Therefore, to maximize the sample size
while balancing velocity coverage against S/N, we adopt a minimum
quasar redshift  of $z_{quasar} =  1.8$.  We select the 1000 quasars with
the highest median S/N in the $r^{\prime}$-band that meet the above
redshift criteria, as returned by the DR4 spectroscopic
query form.  This selection results in a sample with
r$^{\prime}$-band S/N $\ge 17.9$ in each spectrum.

\begin{figure}
  \centering
  {\includegraphics[angle=0,width=80mm]{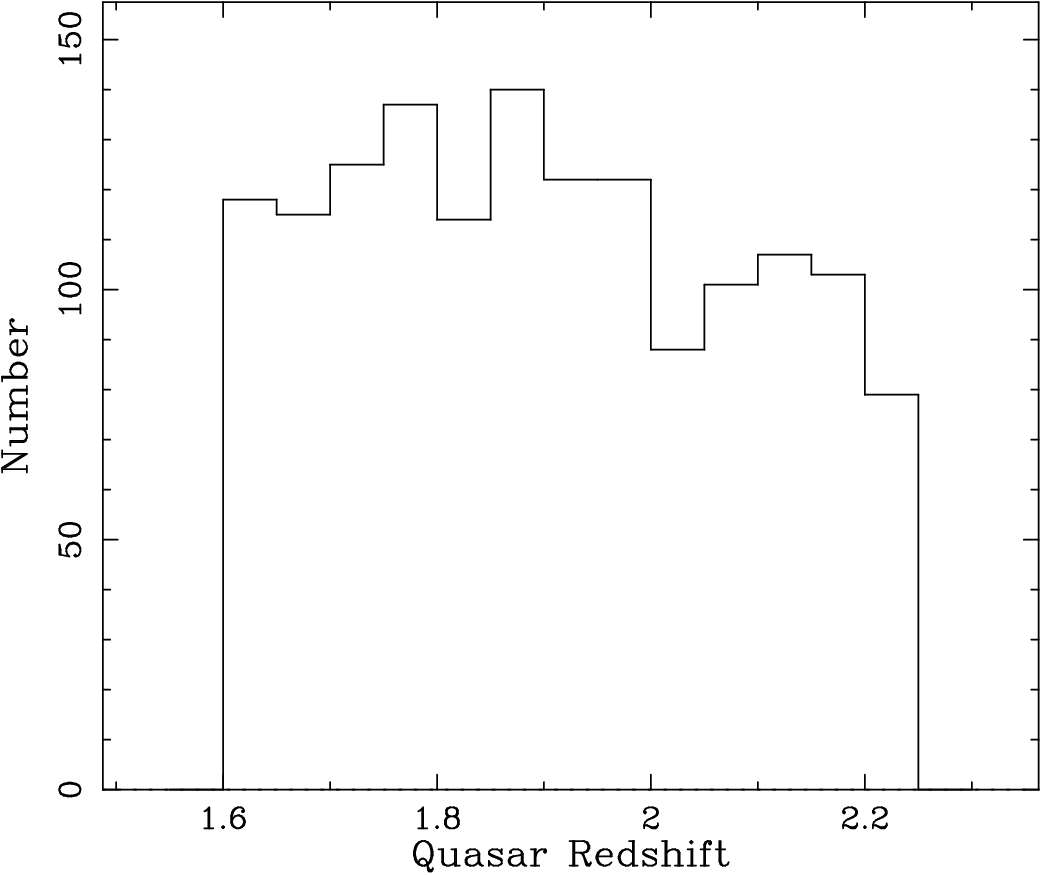}}
  \caption{The redshift distribution of our quasar sample.}  
  \label{Fig:nzq}
\end{figure}

The incidence of absorbers at relatively low quasar-frame velocity  is
of particular interest.  Therefore, in order to increase the sample
size at low-velocity, we select an additional 500 spectra with $1.6
\le z_{quasar} \le 1.8$.  These had r$^{\prime}$-band S/N $ \ge 18.7$
in each spectrum.  The emission redshift distribution of the total
sample of 1500 quasars is shown in Fig. \ref{Fig:nzq}.

\subsection{Velocity zero-point determination}
\label{Sec:mgz}
As quasar broad emission lines are typically blueshifted from the
systemic (centre-of-mass) redshift of the AGN, with the amount of the
shift varying from line to line  (e.g., Tytler \& Fan 1992; Vanden
Berk et al.\ 2001), it is a non-trivial matter  to determine accurate
velocity zero-points in individual spectra.   The SDSS determines
quasar redshifts using an automated algorithm (Stoughton et al., 2002)
to perform a weighted matching of detected emission lines to
empirically-determined wavelengths from  the composite spectrum of
Vanden Berk et al.  These empirical wavelengths, which generally
differ from laboratory wavelengths, are determined assuming that
[\oiii] represents the systemic redshift.  According to Schneider et
al.\ (2002), the statistical error (measured from the scatter from the
individual lines or the height and width of the cross-correlation
function) in this method is less than $\delta_z \simeq 0.01$
($\delta_v \simeq 1000$ \kms) RMS for non-BAL quasars and $\delta_z
\simeq 0.01 \mathrm{-} 0.03$ ($\delta_v \simeq 3000$ \kms) for BAL
quasars.  Additionally, there are also very likely to be systematic
errors that depend on which lines are covered, and thus depend on
redshift in a non-straightforward manner.  Thus, because accurate
velocity zero-points are important to this study, the SDSS-provided
redshifts are insufficient for our purposes.  

It has been shown, however, that the \mgii\ broad emission line has
both a relatively small average offset from [\oiii] ($\left<\Delta
v_{\mathrm{\,Mg\,II}}\right> \sim 0 \pm 100$ \kms) and a relatively
small quasar-to-quasar  scatter in the offset  ($\sigma_{\Delta v}
\la$ a few hundred \kms) compared to other broad emission lines.  The
use of \mgii-derived redshifts should therefore reduce the dispersion
between the measured and true values of the quasar-frame absorption
velocities and thus increase the precision of our measurements.  Even
more important than reducing the scatter in the measured velocities,
however, is to remove any systematic shift by using an accurate and
appropriate average velocity-offset of the \mgii\ emission line
relative to [\oiii].  Tytler \& Fan give a value for this offset of
$\left<\Delta v_{\mathrm{\,Mg\,II}}\right> = +101$ \kms\ with a
dispersion of $\sigma_{\Delta v} = 92$ \kms, where positive velocity
indicates a blueshift with respect to [\oiii].  Richards et al.\
(2002), however, find $\left<\Delta v_{\mathrm{\,Mg\,II}}\right> =
-97$ \kms\ and $\sigma_{\Delta v} = 269$ \kms.  The Richards et al.\
results are derived from a large sample and, like the present study,
using SDSS data, but with quasars at significantly lower redshift than
used here.  McIntosh et al.\ (1999) and Scott et al.\ (2000) provide
data at higher redshift using near-infrared spectra; we made use of
the nineteen high-$z$ values from these two sources and find
$\left<\Delta v_{\mathrm{\,Mg\,II}}\right> = +102$ \kms,  but with a
large dispersion.  Removing one outlier reduces the dispersion and
gives $\left<\Delta v_{\mathrm{\,Mg\,II}}\right> = -1$ \kms.  

Clearly
there is no consensus value for either the offset or the
quasar-to-quasar dispersion in the offset.  Thus, we conduct all of
our analyses using four different techniques for determining quasar
redshifts (and thus velocity zero-points): \mgii-determined redshifts
using \mgii\ zero-point offsets of 0 \kms\ (i.e., no \mgii\
velocity-offset from [\oiii]), $+102$ \kms\  and $-97$ \kms, as well
as the SDSS redshift values.  We adopt the value 
$\sigma_{\Delta v} = 269$ \kms\ from Richards et al.\
as our zero-point uncertainty, which represents the true scatter in
the \mgii-[\oiii] velocity offset convolved with the uncertainty
in the ability to fit the \mgii\ emission in the SDSS data.  Our
interactive fitting (described below) is likely superior to the
automated routine employed by the SDSS pipeline.  However, comparison
of our values to those from the SDSS database indicate the improvement
is in most cases small, and thus the adopted value of 
$\sigma_{\Delta v} = 269$ \kms\ is likely a fair estimate of our
zero-point uncertainty.

\begin{figure}
  \centering
  {\includegraphics[angle=0,width=80mm]{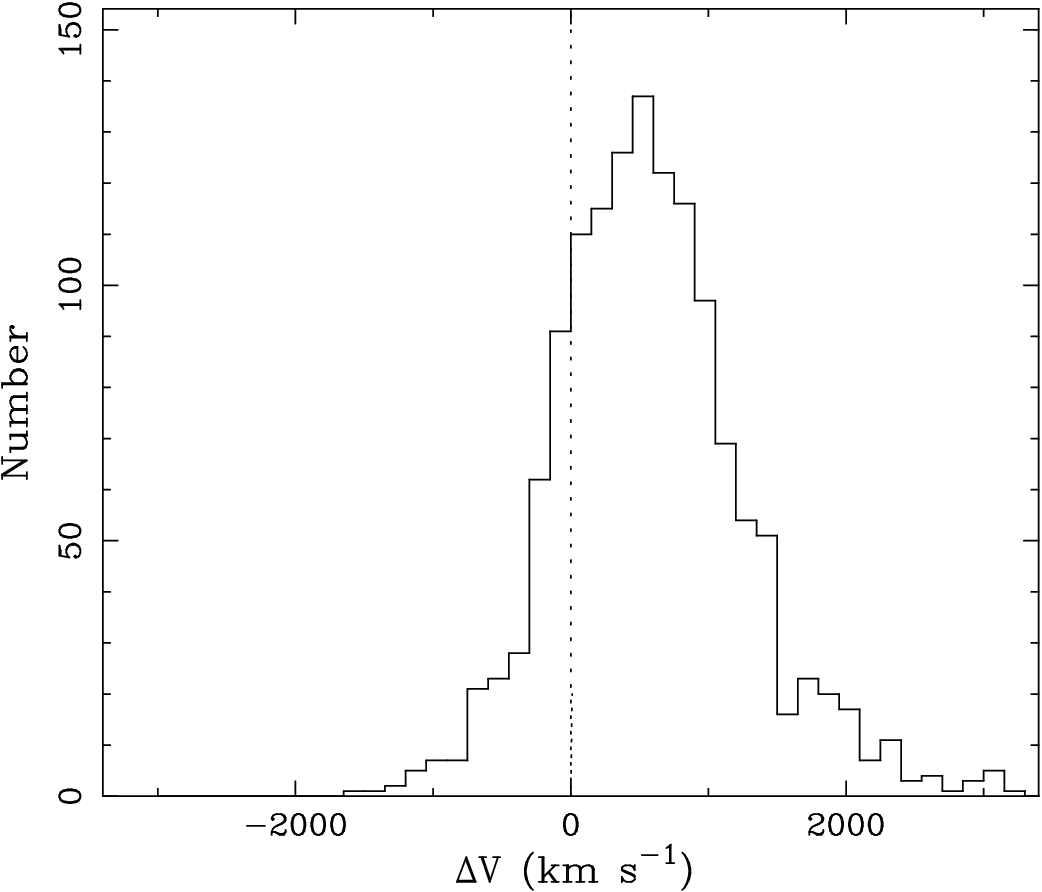}} 
\caption{The velocity zero-point correction distribution of our quasar
  sample for $\left<\Delta v_{\mathrm{\,Mg\,II}}\right> = 0$ \kms.  Positive
  velocities correspond to our \mgii-corrected redshift being larger
  than the SDSS redshift.  The distribution has a mean of $+582$
  \kms\ and a standard deviation of 684 \kms.}  
\label{Fig:dz}
\end{figure}

To compute the \mgii-determined redshifts, we fitted a Gaussian to the
\mgii\ emission line for each quasar in our sample, assuming the given
velocity zero-point offset.  The fits made use of only the central
peak of the emission line in order to avoid effects from
line-asymmetries.  In practice, the extent of the fitted region
varied, depending on the emission line profile.  The precision with
which we were able to determine the peak wavelength also depended on
the profile.  The formal (i.e., fitting) errors were typically $<$ 10
\kms.  While the actual \mgii\ zero-point uncertainties  are
necessarily larger than this, they are in most cases always well below
the dispersion as described above.   Such fits were possible in 92 per
cent of the sample; while the \mgii\ emission in the remaining 8 per
cent was too weak and/or noisy (e.g., due to poorly subtracted
night-sky lines) to determine an accurate fit.  For these systems, we
applied an assumed zero-point correction equal to the average
correction of the \mgii-fitted sample.  The distribution of velocity
zero-point differences (compared to the SDSS values) for the
re-determined redshifts is shown in Fig. \ref{Fig:dz} for no \mgii\
velocity zero-point offset.  Positive velocities correspond to our
\mgii-derived redshift being larger than the SDSS redshift.  The
distribution has a median value of $+545$ \kms, a mean of $+582$ \kms\
and a standard deviation of 684 \kms.  The velocity zero-point 
difference distribution for the SDSS-provided redshifts and those determined
from the \mgii\ emission fits in the SDSS database is almost
identical to Fig. \ref{Fig:dz}, except for a large spike at $+1500$ \kms\ as
the SDSS line-fitting algorithm does not allow shifts larger than this
value.

\subsection{Identification and measurement of C\,{\sevensize\bf IV} absorbers}
We fit pseudo-continua (including the broad emission lines)  to each
spectrum and identify \civ\ absorber candidates using software
described in Nestor, Turnshek \& Rao (2005), updated and adjusted to
apply to \civ.  Each continuum-fit was visually inspected and, when
necessary, interactively re-fit.  Approximately 2 per cent of the
quasars in our sample were either misclassified stars or exhibited BAL
absorption strong enough that there was little unabsorbed flux in the
spectral regions of interest and accurate continuum-fitting proved
impossible.  These objects were removed from the sample.

The software used to identify \civ\ candidates is very sensitive to
narrow absorption; therefore we are confident that all narrow
absorbers above our significance level threshold ($\geq 5\sigma$ for the 
$\lambda1548$ line and $\geq 3\sigma$ for $\lambda1551$, see Nestor et  al.\
2005) were found.  Each candidate was visually inspected to  remove
redundant and false detections.  Care was taken to avoid
misclassifying the low-ionization pair \hbox{C\,{\sc i}} $\lambda1277$
and \hbox{O\,{\sc i}} $\lambda1302$, which has a similar separation
to the \civ\ doublet, as high-velocity ($v \ga 48000$ \kms) \civ\
absorption.  Furthermore, the Ly$\alpha$ and \nv\ blended emission
lines occasionally led to difficulties for our continuum-fitting
software.  Thus, while we did not exclude these regions when detecting
and measuring absorbers, we conservatively restrict the analyses in
this paper to velocities $v \le 60000$ \kms\ in order to avoid
potential biases.

The large majority of the \civ\ absorbers that we identified in our
sample were straightforward to measure.  However, there are also many
systems exhibiting broad, resolved profiles.  Furthermore, as \civ\
absorption is relatively common in quasar spectra, blending of systems
at similar redshifts is occasionally a problem in the medium
resolution SDSS spectra, particularly at low-velocity ($z_{abs}
\approx z_{quasar}$) where there is an excess of absorbers (see
below).  While highly accurate rest equivalent width ($W_0$) and FWHM
measurements are not crucial for the purposes of this paper, proper
{\it counting} of systems is very important.  It is typical in studies
of intervening absorbers to designate a minimum velocity separation
for which two absorbers are counted separately.  However, series of
systems at low-velocity can be blended across large velocity ranges,
and narrow systems can be partially blended with broad systems, making
the effective separation for which it is possible to discern separate
systems widely variable.  Therefore, we took the following approach,
that will be described in greater detail in Rodriguez Hidalgo, Hamann
\& Nestor (in prep.)  To each system we fit a Gaussian pair having the
correct \civ\ doublet separation, the same FWHM, and a
$\lambda1548:\lambda1551$ doublet ratio constrained to the range $1:1$
to $2:1$.  When absorbers appeared to have distinct but blended
components, we simultaneously fit multiple pairs to the entire blended
profile, with each component fit with a single Gaussian doublet pair
and considered as a distinct absorption system.  In order to decide
the number of components to fit, we always used the minimum number
justifiable by the data; that is, we only added an additional
component when there was clear evidence for a separate absorption
feature (member of a doublet).  In some cases, broad components were
not well described by a single Gaussian pair, but the data did not
justify deblending into multiple features according to our counting
rules.  Nonetheless, a  single Gaussian pair per component always
provided a reasonable estimate of both $W_0$ and FWHM for the system
in question.  Since the problematic systems are exclusively broad,
they do not enter into the analysis described in this paper.

\section{Analysis}
\label{Sec:analysis}
\subsection{Defining the narrow C\,{\sevensize\bf IV} sample}
\label{Sec:abssample}
For the present study, we want to exclude the more obvious outflows
lines, the BALs and mini-BALs (\S\ref{Sec:intro}). We therefore
removed from our sample all absorbers with measured FWHM $> 600$ \kms.
We also removed all absorbers blended with absorbers having FWHM $>
600$ \kms\ as their identification can often be ambiguous. This 
helps reduce uncertainties associated with deblending
and counting but, in the end, has little impact on our results
because the number of such  systems is small ($<$1 per cent of the sample).  
Finally, several
spectra in the sample contain extremely strong BAL absorption which
removes flux from a large swathe of continuum.  We computed both the
BALnicity Index (BI; Weymann et al.\ 1991) and the Absorption Index
(AI;  Hall et al.\ 2002) to quantify these absorbers (see Rodriguez
Hidalgo  et al.\ for more discussion).  After visual inspection, it
was decided to remove spectra with AI$>3120$ and/or BI$>1250$, which
totaled 62 quasars.  We note that this is {\it not} the equivalent of
removing BAL quasars; rather, we remove only those with little
`usable' unabsorbed continuum (although, for the BAL quasars  kept in
the sample, we do not include narrow \civ\ systems that are blended
with BALs.)

\begin{figure}
  \centering
  {\includegraphics[angle=0,width=80mm]{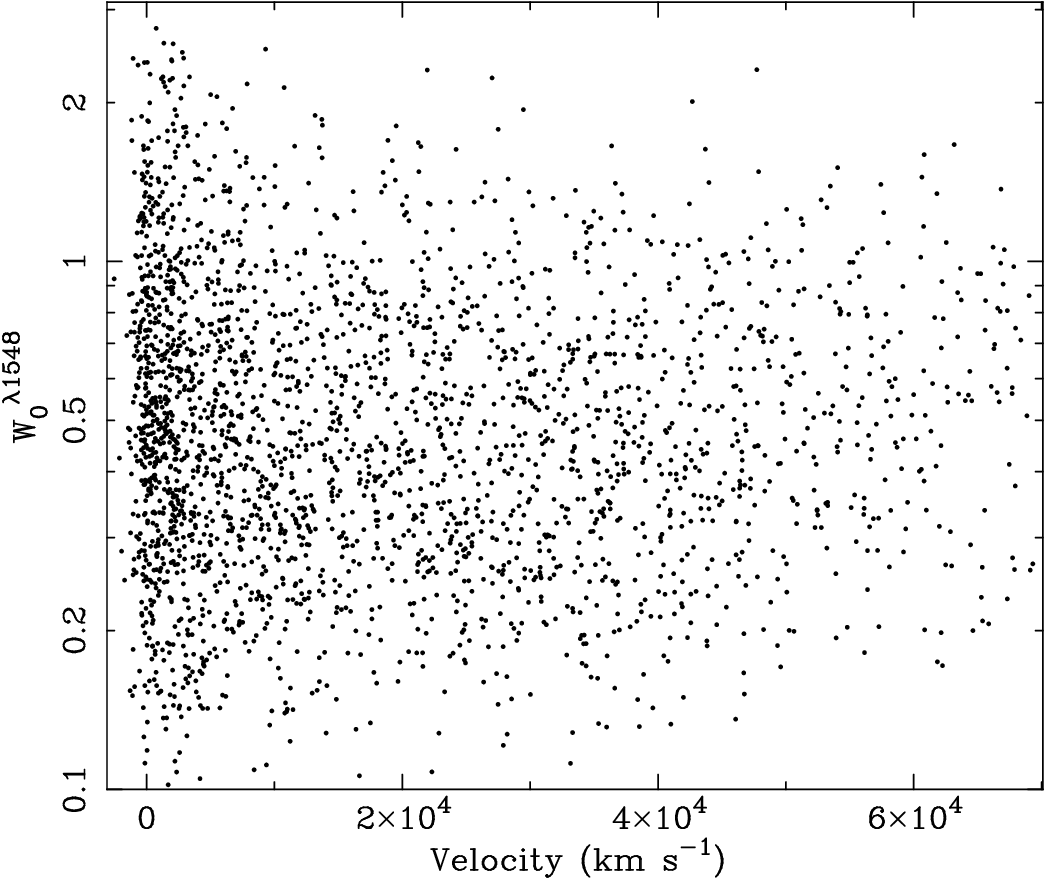}}
  \caption{Scatter-plot of \W\ versus quasar-frame velocity for our
    narrow (FWHM$\le$ 600 \kms) absorber sample.  Velocities shown are for
    the \mgii-corrected redshifts assuming no
    \mgii\ velocity zero-point offset.}  \label{Fig:wv}
\end{figure}

The final sample includes 1409 quasars, which provide a catalog of
2566 narrow \civ\ absorbers with $0.1 < $ \W\ $< 2.8$ \AA\ and $-3000
< v_{abs} < 70000$ \kms, 2009 of which  have \W\ $> 0.3$\AA.  We note
that \W\ and FWHM are correlated, so the removal of FWHM $> 600$ \kms\
systems biases our catalog against very strong systems.  Figure
\ref{Fig:wv} shows a scatter-plot  of \W\ versus quasar-frame velocity
for the final narrow \civ\ catalog.

\subsection{Computing the completeness correction}

\begin{figure}
  \centering
  {\includegraphics[angle=0,width=80mm]{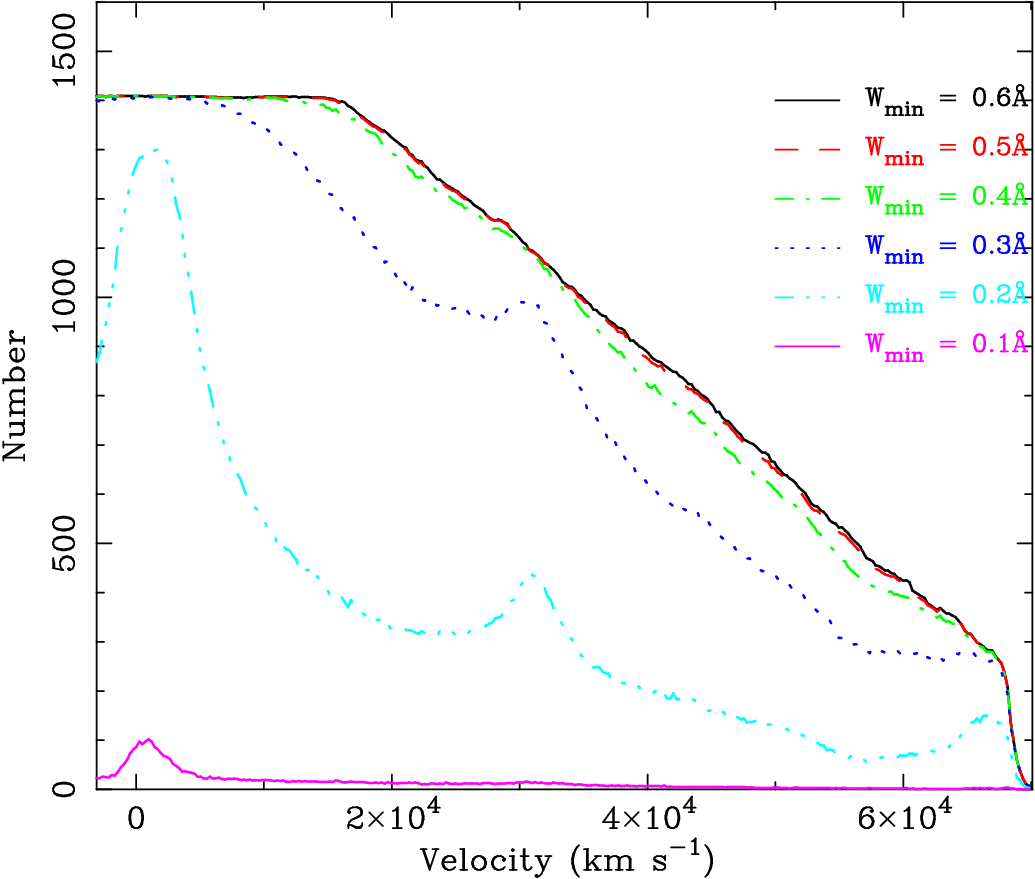}}
\caption{The
  number of usable spectra as a function of velocity corresponding to
  \W\ values of 0.1\AA, 0.2\AA, 0.3\AA, 0.4\AA, 0.5\AA\ and 0.6\AA,
  using \mgii-corrected redshifts assuming no \mgii\ velocity-offset.
  The 0.5\AA\ and 0.6\AA\ curves are nearly identical -- above
  \W$\simeq0.5$\AA\ our sample is complete in \W\ at all velocities.}
  \label{Fig:vpath}
\end{figure}

Our data set is complete in neither \W\ or velocity-space.  It is
therefore necessary to compute a two-dimensional completeness
correction.  We follow a procedure similar to that described in Nestor
et al.\ (2005), whereby we determine the minimum \W, $W_{min}$,
detectable at our imposed significance-level threshold at every pixel
and half-pixel in each spectrum.  We then construct an array on a grid
of velocity and \W\ steps with values corresponding to the number of
spectra for which a \civ\ $\lambda1548$ line of the given \W\ could
have been detected at the given velocity, accounting for the removed
regions of spectra having absorbers with FWHM $> 600$ \kms.  This
`completeness array' can then be used to correct any computed
incidence of absorbers in both \W\ and velocity space.  In
Fig. \ref{Fig:vpath} we show the number of usable spectra as a
function of velocity for different minimum \W\ values.  The 0.5\AA\
and 0.6\AA\ curves are nearly identical -- above \W$\simeq0.5$\AA\ our
sample is complete in \W\ at all velocities.
Although the minimum detectable \W\ depends, in principle, on the
absorption FWHM, in practice it is computed assuming unresolved lines.
However, any detection bias arising from this would only be relevant
for lines with \W$ \simeq W_{min}$ that are also resolved in velocity.
However, lines weaker than $\approx 0.5$\AA\ (the strength above which
our survey is complete) are generally unresolved in the SDSS spectra.

\subsection{Results}
\label{Sec:results}
\subsubsection{The quasar-frame velocity distribution}
\label{Sec:veldist}
\begin{figure}
  \centering
  {\includegraphics[angle=0,width=80mm]{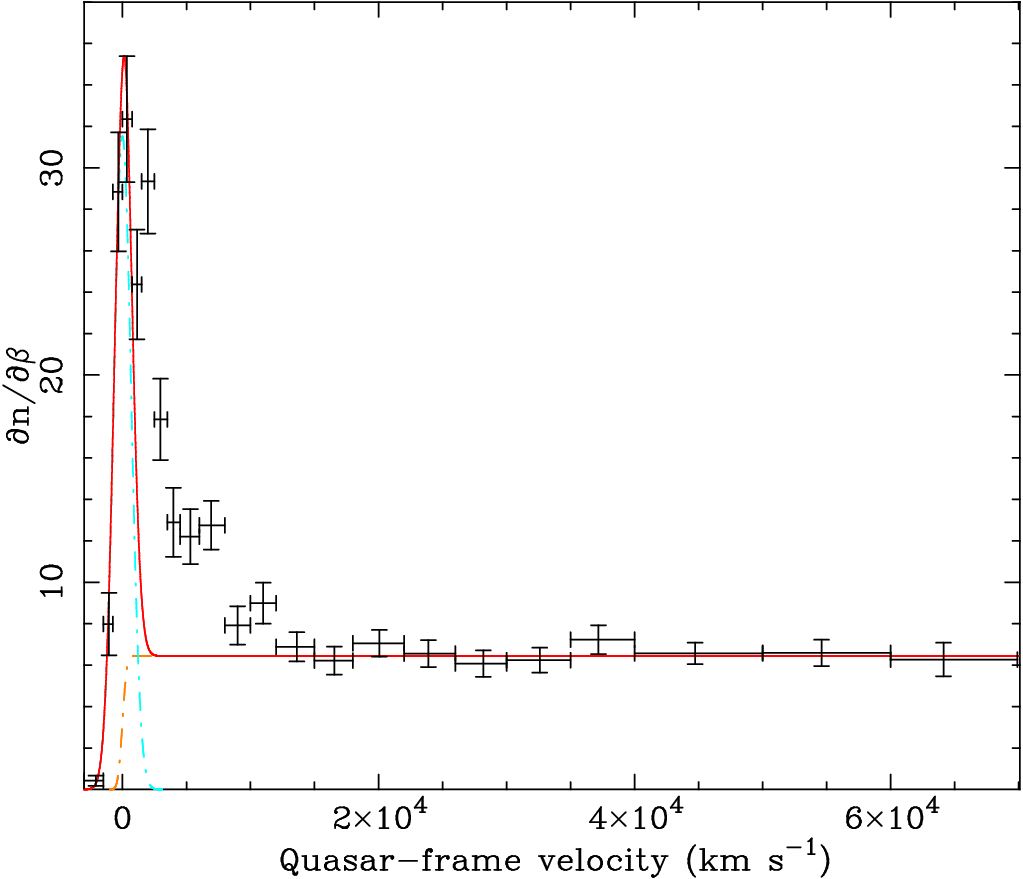}} 
\caption{The incidence of \civ\ absorbers versus quasar-frame velocity
for systems with \W\ $\ge 0.3$\AA, using \mgii-corrected redshifts and
assuming no \mgii\ velocity-offset.  Horizontal bars indicate the
velocity bins and vertical bars the 1 $\sigma$ uncertainty from
counting statistics.  The curves are maximum-likelihood fits to the
data in the intervals $v<0$ and 40000 \kms\ $< v < $ 60000 \kms,
representing the `environmental' and `intervening' components,
respectively (see text).}
  \label{Fig:dndv1}
\end{figure}

Using the completeness array together with the absorber catalog, we
can compute the incidence of absorbers as a function of velocity,
$\frac{\partial n}{\partial \beta}$, where $\beta= v/c$.  The
magnitude of $\frac{\partial n}{\partial \beta}$ represents the
average number of absorbers per \kms\ times the speed of light.  Thus
the average number of absorbers per spectrum can be found by
integrating $\frac{\partial n}{\partial \beta}$ over the relevant
velocity range:
\begin{equation}
\label{Eqn:dndv}
   \left< n \right> = \frac{1}{c}\int_{V_{min}}^{V_{max}}
   \frac{\partial n}{\partial \beta}\,dv.
\end{equation}
We show $\frac{\partial n}{\partial \beta}$ for \W\ $\ge 0.3$\AA,
using \mgii-corrected redshifts and assuming no \mgii\
velocity-offset, in Fig. \ref{Fig:dndv1}.
A large excess at $v \approx 0$ (i.e., $z_{abs} \simeq z_{quasar}$) is
clearly detected.  The observed distribution includes contributions
from: (i) cosmologically-`intervening' systems; (ii) `environmental'
absorption which arises in either the ISM of the AGN host galaxy or
the IGM of the galaxy's group/cluster; and (iii) `outflow' systems
that are ejected from the central AGN (e.g., accretion-disk outflows).
We model the intervening contribution to the distribution with a step
function that breaks  at $v=0$, convolved with the redshift
uncertainty (modeled as a Gaussian with $\sigma_z = 269$ \kms). We
model the environmental contribution with a Gaussian centred at
$v=0$.  To obtain the best-fit parameters, we employ a maximum
likelihood fit to the unbinned data (absorber catalog plus
completeness array), using only data with $v<0$ and 40000 \kms\ $< v <
60000$ \kms\ to constrain the fit, where the outflow fraction is {\it
assumed} to be insignificant at these velocities.  The physical
velocity dispersion of the AGN-host's local environment can be
estimated from the environmental-component by deconvolving the
Gaussian fit with the assumed redshift uncertainty (269 \kms) and,
following Weymann et al., divide by $\sqrt{2}$ to account for the fact
that what we measure is the difference between the velocity of the
AGN-host within the environment and that of the other halos (although
the appropriateness of this last step is limited by the presence of
absorption in the AGN host galaxy, which is not virialized in the
local environment).   

The parameters describing the fits are shown in Table \ref{Tab:fits}
for each zero-point determination method (\S\ref{Sec:mgz}).  The first
column indicates the velocity zero-point offset (for the
\mgii-determined redshifts cases), and the parameters are defined by:
\begin{equation}
   \frac{\partial n_{env}}{\partial \beta} = N_{env} \times \exp(-v^2/2\sigma^2_{env}),
\end{equation}
\begin{equation}
   \sqrt{2}\times \sigma^{\prime}_{env} = \sqrt{\sigma^2_{env} - (269\ \mathrm{\kms})^2},
\end{equation}
and
\begin{equation}
   \frac{\partial n_{inter}}{\partial \beta} = N_{int}.
\end{equation}
\begin{table}
  \caption{Environmental and intervening fit-parameters.}
\begin{center}
\label{Tab:fits}
  \begin{tabular}{ccccc}
$\left<\Delta v_{\mathrm{\,Mg\,II}}\right>$   &           & $\sigma_{env}$  &    $\sigma^{\prime}_{env}$ &\\  
(\kms)                           & $N_{env}$  & (\kms)         & (\kms)                   & $N_{inter}$\\
\hline
0                                &     31.5  & 681            & 442                      & 6.4 \\  
$+102$                           & 26.8      & 665            & 430                       & 6.4    \\  
$-97$                            & 35.4      & 694            & 452                       & 6.4 \\ 
SDSS                             & 25.9      & 1319           & 913                      & 6.5 \\  
\hline
\end{tabular}
\end{center}
\medskip

The $\sigma_{env}$ values are the Gaussian widths before and after
deconvolution and the $N$ values are the normalization of the fits --
i.e., the incidence for intervening systems and the incidence of
environmental systems at $v=0$.
\end{table}

Our assumption that {\it none} of the absorbers at $v > 40000$ \kms\
or $v<0$ arise in outflows means that our derived outflow fractions
(see below) are strictly speaking lower limits at all velocities.  For
example, Richards et al.\ claim that as many as 36 per cent of \civ\
absorbers at $5000 < v < 55000$ \kms\ form in outflows, while Misawa
et al.\ (2007)  claim that the fraction of outflow lines at these
velocities is $\simeq$ 10--17 per cent.  However, it is unclear what
the contribution of very high-$v$ systems (i.e., $v > 40000$ \kms) to
the Richards result is, and the Misawa et al.\ sample contains no
systems at $v > 40000$ \kms\ that are classified as a reliable
narrow-line outflow candidates.  Thus, while we cannot directly test
our high velocity assumption, it is likely that we are only slightly,
if at all, overestimating the contribution of intervening systems.

The assumption that all $v<0$ absorbers are environmental is
potentially less reliable, however.  It will be an appropriate
assumption if there are no `in-falling' systems intrinsic to the AGN
flow and all of the individual quasar velocity zero-points are
accurate (or that there are no outflowing systems with velocities that
are small relative to the velocity zero-point dispersion).  However,
our zero-points are inaccurate with an (assumed) RMS of $\sim$ 270
\kms, and the systematic velocity-offset of \mgii\  from the true
velocity zero-point is not well constrained (see \S\ref{Sec:mgz}).
Thus it is possible that our environmental fits are also accounting
for some low-velocity outflow systems.  We investigate this issue
below.

\begin{figure}
  \centering
  {\includegraphics[angle=0,width=80mm]{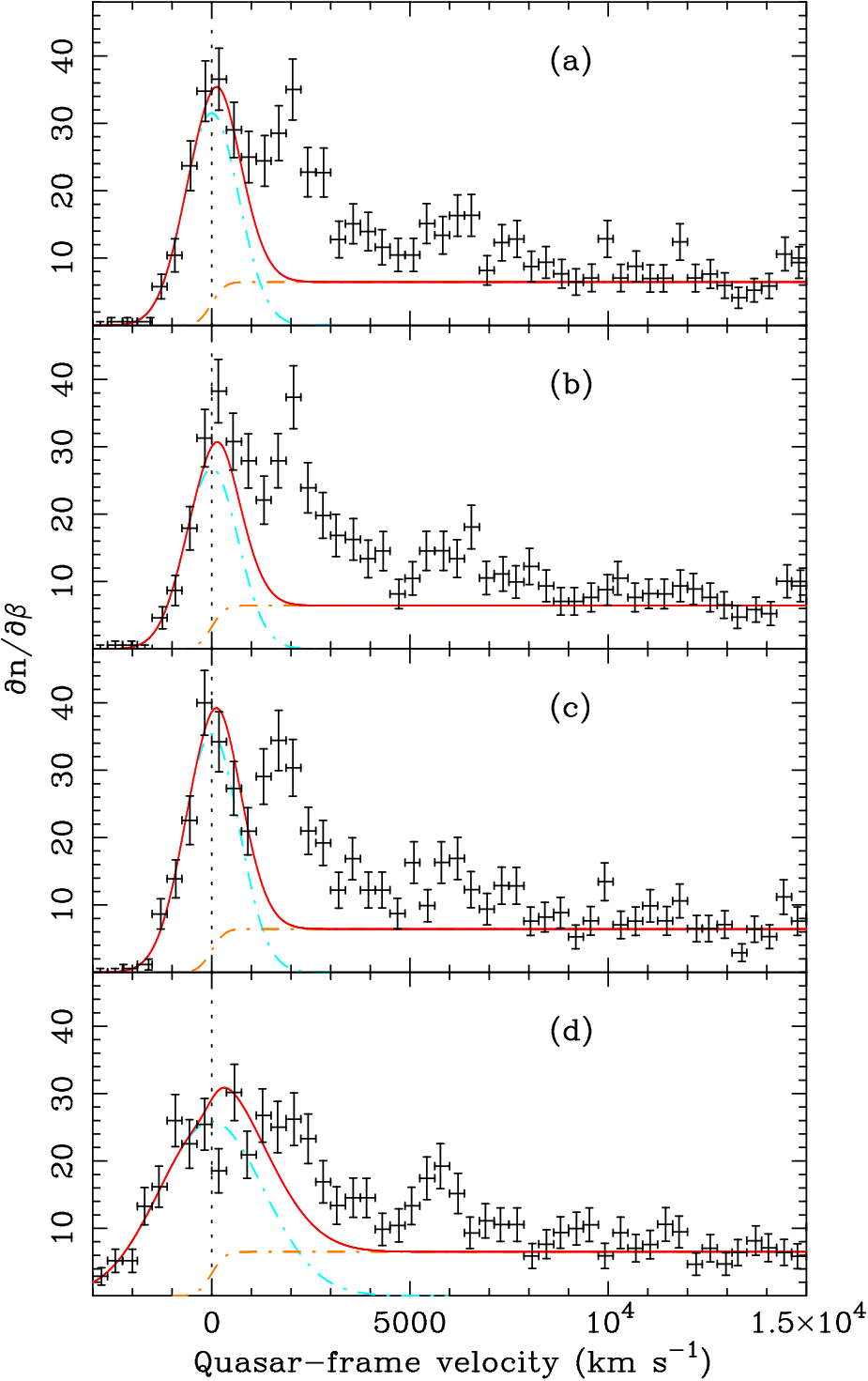}} 
\caption{Same as Fig. \ref{Fig:dndv1}, employing  \mgii-determined
redshifts with (a) no \mgii\ velocity zero-point offset, (b)
$\left<\Delta v_{\mathrm{\,Mg\,II}}\right> = +102$ \kms, (c)
$\left<\Delta v_{\mathrm{\,Mg\,II}}\right> = -97$ \kms, as well as (d)
the SDSS-provided redshifts.   The width of the
environmental-component determined from SDSS redshifts is unphysically
large, and that determined assuming $\left<\Delta
v_{\mathrm{\,Mg\,II}}\right> = -97$ \kms\ gives a poor fit to the $v \sim 0$
data.}  
\label{Fig:dndv2}
\end{figure}

The uncertainty in the appropriate velocity zero-point has a direct
effect on the measured velocity-space distribution of absorbers.  In
order to determine the magnitude of this potential bias and
investigate its affect on the computed outflow fraction, we computed
$\frac{\partial n}{\partial \beta}$ as well as the maximum-likelihood
fits using all four of the quasar redshift determination methods
discussed in \S\ref{Sec:mgz}.  Fig. \ref{Fig:dndv2} shows the
low-velocity region of $\frac{\partial n}{\partial \beta}$ computed
using:  \mgii-determined redshifts with (a) no \mgii\ velocity
zero-point offset (i.e., $\left<\Delta v_{\mathrm{\,Mg\,II}}\right> =
0$), (b) $\left<\Delta v_{\mathrm{\,Mg\,II}}\right> = +102$ \kms, (c)
$\left<\Delta v_{\mathrm{\,Mg\,II}}\right> = -97$ \kms, as well as (d)
SDSS redshifts.  The curves
represent the data well in the fit-regions, for all of the offsets
considered.  However, the
distribution determined assuming $\left<\Delta
v_{\mathrm{\,Mg\,II}}\right> = -97$ \kms\ 
overpredicts the incidence at $v \ga 0$.  We note that, if the
velocity zero-point is (on average) properly determined, the incidence
for $v \ga 0$ should be larger than that for $v \la 0$, since the $v
\ga 0$ data contains, in principle, environmental, intervening, and
outflowing \civ\ systems, and we assume the environmental-component is
reasonably symmetric about the true velocity zero-point.
We obtain a value of $\sigma^{\prime}_{env} = 913$
\kms\ using the SDSS redshifts, which is unphysically large unless the
quasar environments are predominantly in large clusters (e.g., Becker
et al.\ 2007).  However, Outram et al.\ (2003) show that the 
clustering amplitude of quasars at $z \sim 1.4$ is similar to that 
of present-day galaxies.  The \mgii-determined redshifts
lead to  values of $430 \la \sigma^{\prime}_{env} \la 450$ \kms\ for
the dispersion, which is comparable to the usual division between
large groups and poor clusters (e.g., Mulchaey, 2000).  

Considering the quality of the fits, we conclude that, while we do not
explicitly reject any of the \mgii-corrected redshift choices, an average
\mgii\ velocity zero-point offset of $\simeq 0$ to $+100$ \kms\ is
most appropriate for our sample.  Except when explicitly stated, we
will use $\left<\Delta v_{\mathrm{\,Mg\,II}}\right> = 0$ for the
subsequent analyses.  We also note that we computed $\frac{\partial
n}{\partial \beta}$ for ranges of \W, but found no significant trends.  

\subsubsection{The outflow fraction and incidence of outflowing absorbers}
\begin{figure}
  \centering
  {\includegraphics[angle=0,width=80mm]{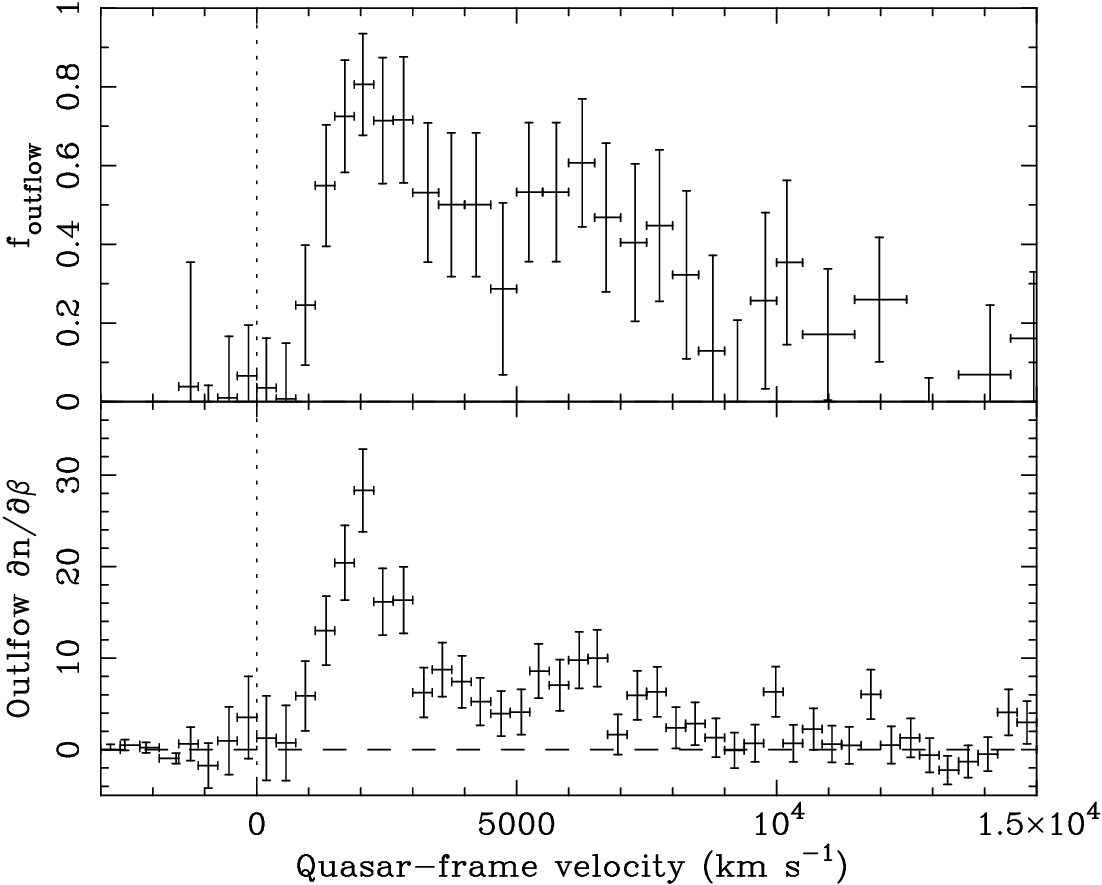}} 
  \caption{Top: The excess of the data over
the sum of the two fits, divided by the data,
which represents the minimum outflow fraction.  Bottom:
The data minus our fits to the environmental and
    intervening systems, representing $\frac{\partial n}{\partial
      \beta}$ for the ejected (i.e., outflow) component only.}
  \label{Fig:dndvi}
\end{figure}

The top panel of Fig. \ref{Fig:dndvi} shows the data minus the sum of
the two fits, divided by the data, which represents the minimum
outflow fraction, $f_{outflow}$ (which is considered a minimum for the
reasons described above).  We find that $f_{outflow}$ increases
strongly from $v=0$ to $v \simeq +2000$ \kms, where it peaks with
$f_{outflow} \simeq 0.81 \pm 0.13$, and then decreases slowly out to
$v \sim +12000$ \kms.  Over the range where there is significant
evidence for a non-zero outflow fraction, $v \simeq +750$ \kms\ to $v
\approx +12000$ \kms, we find $\left<f_{outflow}\right> =
0.43\pm0.06$.  Over narrower ranges near the peak, we find
$\left<f_{outflow}\right> = 0.57\pm0.10$ for +1250 \kms\ $< v < $
+6750 \kms, and $f_{outflow} \simeq 0.72 \pm 0.07$ for $v \simeq
+1250$ \kms\ to $v \approx +3000$ \kms.  The outflow fraction
decreases below $v \approx +2000$ and disappears below $v \approx
+750$, apparently indicating an effective minimum projected 
ejection velocity for narrow \civ\ systems.  
It is also possible that systematics such as a
strong proximity effect for the intervening absorbers or limitations
in our ability to properly count systems contribute to this decrease.
We discuss these and other possible biases in \S\ref{Sec:syst}.  

In the bottom panel of Fig. \ref{Fig:dndvi} we show the data
after subtraction of the sum of the fits (using $\left<\Delta
v_{\mathrm{\,Mg\,II}}\right> = 0$), which represents $\frac{\partial
n}{\partial \beta}$ for the outflow-component only.
Using $\left<\Delta v_{\mathrm{\,Mg\,II}}\right> = -97$ \kms\ (i.e.,
\mgii\ emission redshifted from the quasar rest-frame) or
SDSS-determined redshifts leaves, respectively, marginally- and 
strongly-statistically significant negative, and thus unphysical,
residual incidence  for the outflow-component at some velocities.
Using $\left<\Delta v_{\mathrm{\,Mg\,II}}\right> = 0$ (i.e., \mgii\
emission unshifted from the quasar rest-frame), the incidence of
systems is consistent with zero at low positive velocities: in this
case, our assumptions that no absorbers arising in outflows have
measured velocities with $v < 0$  is likely valid.  Attempts to fit the
entire span of velocities (i.e., including $0<v<40000$\,\kms, 
see \S\ref{Sec:veldist}) with, 
e.g., a power-law or exponential for $v > 0$, failed since, as is
clear from the bottom panel of Fig. \ref{Fig:dndvi},
these were not appropriate descriptions of the outflowing component.
Considering the
results employing \mgii-corrected quasar redshifts, we find that the
incidence of narrow \civ\ absorbers with \W\ $> 0.3$ \AA\ presumably
forming in accretion-disk outflows peaks at $v \simeq 2000$ \kms\ and
exhibits a wing that extends out to at least $v \simeq 9000$ \kms, for
all $\left<\Delta v_{\mathrm{\,Mg\,II}}\right>$ values considered.   

\subsection{Possible systematics}
\label{Sec:syst}
In this section we discuss potential biases introduced by the method
we employ to `count' systems as well as the assumptions behind the
modeling of the distribution of cosmologically-intervening absorbers.  

As narrow \civ\ absorbers cluster at low ($\la 1000$ \kms) 
quasar-frame velocities (Nestor,
Hamann \& Rodriguez Hidalgo, 2007), removing regions with broad
absorption is more likely to also remove narrow absorbers, compared to
randomly selected regions at similar velocity.  Since broad absorbers
are more common at low positive velocity, it is a concern that this
bias may affect the measured distribution of narrow systems.
Furthermore, this region of velocity space exhibits (in principle) all
three of our categories of absorbers, thus increasing the chances for
blending of outflow and non-outflow systems to occur.  Therefore, the
apparent dearth of narrow outflow-component systems at $v \la 2000$
\kms\ (Fig. \ref{Fig:dndvi}) could conceivably be in part a byproduct
of our removal of broad systems and absorbers blended with broad
systems.  To test this, we re-computed $\frac{\partial n}{\partial
\beta}$ without rejecting {\it any} absorbers based on FWHM or
blending with broad systems.  However, the only qualitative difference
in our results was an extension of the high-velocity tail of
outflowing systems out to $v \approx 25000$ \kms, indicating that
broad (FWHM$> 600$ \kms) absorbers have a larger extent in
ejection-velocity than do narrow systems.  This will be discussed in
further detail in Rodriguez Hidalgo, Hamann \& Nestor (in prep.)  

Alternatively, over-zealous de-blending could, conceivably, influence
both the shape and magnitude of $\frac{\partial n}{\partial \beta}$.
To investigate this concern, we combined all absorbers with velocity
differences less than the sum of the two HWHM values plus 150 \kms\
(i.e., the SDSS resolution) into a single absorber to be counted once.
We then recomputed $\frac{\partial n}{\partial \beta}$ and the fits
with this reduced catalog.  This resulted in a slight decrease in the
outflow fraction for bins where it was non-zero, and virtually no
change in the shape or range over which $f_{outflow} > 0$ is
significant.  Even when we took the extreme step of  combining all
absorbers within 500 \kms\ plus the sum of the two HWHM values, the
shape of $\frac{\partial n}{\partial \beta}$ for outflow-only systems
remained the same, although the magnitude was reduced by $\simeq 20$ -
25 per cent.  Thus, we believe that any inconsistencies in our
counting method is at a level small enough to be safely ignored.

Another concern is that cosmologically-intervening systems may not be
uniformly distributed in quasar-frame velocity for our quasar sample.
For example, it is known that at high-$z$ the incidence of intervening
\civ\ absorbers decreases with increasing redshift (Sargent,
Boksenberg \& Steidel, 1988; Misawa et al., 2002).  Larger redshift
correlates with smaller velocity in
Figs. \ref{Fig:dndv1}-\ref{Fig:dndvi}.  According to Monier et al.\
(2006), the incidence of \W\ $\ge 0.3$\AA\ systems show little
evolution over the range $1.46 < z < 2.25$.  Misawa et al., however,
claim a $\simeq 50$ per cent decrease over the same range.  While the
decrease in $f_{outflow}$ is too abrupt to be entirely caused by
redshift effects, we nonetheless investigated the maximum possible
magnitude of this effect by running 50 Monte Carlo simulations of
intervening systems in our data.  We randomly
distributed absorbers in the spectra of our sample in redshift-space 
using the $\frac{\partial n}{\partial
z}$ parameterization from Misawa et al., and converted the resulting
distributions into velocity-space to determine the modelled 
$\frac{\partial n}{\partial \beta}$.  This resulted in $\la 10$ per
cent decrease in $\frac{\partial n}{\partial \beta}$ from $v \approx
45000$ \kms\ to $v=0$, or a difference of $\la 0.7$ in $\frac{\partial
n}{\partial \beta}$.

Potentially more important is the possibility of a strong proximity
effect, similar to that seen for Lyman-$\alpha$ forest lines, causing
a decrease in $\frac{\partial n}{\partial \beta}$ for the intervening
systems.  While the magnitude of any proximity effect likely depends
on the line-strengths being considered, \civ\ has a much higher
ionization energy than does \hi\ (64.5 eV compared to 13.6 eV) and
thus any \civ\ proximity effect should be significantly weaker than
that for \hi.  Thus, we ran 50 additional Monte Carlo simulations of
intervening systems using the proximity effect results for the
Lyman-$\alpha$ forest from Scott et al.\ (2000), considering this a
strong upper-limit to any \civ\ proximity effect.  Doing so, we find a
linear decrease in $\frac{\partial n}{\partial \beta}$ from no
difference at $v = 2000$ \kms\ to a 50 per cent deficit of systems at
$v=0$.  This corresponds to a difference in $\frac{\partial
n}{\partial \beta}$ at $v=0$ of $\simeq 3.5$.  Thus, a \civ\ proximity
effect as strong as what is claimed for Lyman-$\alpha$ would only
marginally change the results described above, and then only at $v
\sim 0$.

Finally, we also considered the effect of an underestimation of the
velocity zero-point dispersion.  A larger dispersion would cause more
low-velocity intervening systems to scatter to $v<0$.  We repeated the
modeling of the environmental and intervening components doubling the
velocity zero-point dispersion and found no qualitative changes to our
results.

We are limited in our ability to determine true individual velocity
zero-points, to deblend (`count') absorbers into physically-distinct
systems with complete accuracy, and to know the true distributions of
the non-outflowing absorber populations.  These systematics almost
certainly affect our results to some degree.  Nonetheless, the maximum
estimates of the effects from all of the systematics that we
considered are relatively small and have no significant impact on our
qualitative results.

\section{Discussion}
\label{Sec:disc}
\subsection{Properties of the outflowing absorbers}
We find a significant detection of narrow \civ\ absorption over that
which is expected from cosmologically-intervening and host
galaxy-environment contributions, putatively occurring in accretion
disk winds, in the quasar-frame velocity range 750 \kms\ $\la v \la
12000$ \kms.  Using eqn. 1 together with the $\frac{\partial
n}{\partial \beta}$ values for the outflow-component
(Fig. \ref{Fig:dndvi}), we can compute the average number of
outflowing absorbers per quasar.  We find $\left<n_{outflow}\right> =
0.25 \pm 0.02$ using $\left<\Delta v_{\mathrm{\,Mg\,II}}\right> = 0$.
The value depends only slightly on the choice of $\left<\Delta
v_{\mathrm{\,Mg\,II}}\right>$ and the limits of integration.  

While it is likely that some \W\ $\ge 0.3$\AA\ low-velocity ($\la 750$
\kms) narrow \civ\ absorbers arise in outflows, the various possible
systematic effects discussed above are unable to completely, or even
mostly, explain the turnover in $\frac{\partial n}{\partial \beta}$
for outflow-only absorbers below $v \la 2000$ \kms\ -- it appears
rather to be an indication of a minimum projected line-of-sight
ejection velocity for narrow \civ\ absorption systems in AGN outflows.
This empirical phenomenon can help to constrain quasar outflow models.
For example, the deficit of \civ\ outflow-systems at $v\la 2000$ \kms\
might result from  a tendency to `over-ionize' the low-velocity gas,
e.g., if that gas is typically nearer the ionizing continuum source
than the high speed absorbers. This can occur in any model of the
acceleration (radiation pressure or magneto-centrifugal forces), but
over-ionization at low velocities has a direct physical explanation if
there is radiative driving. In particular,  highly ionized gas is more
transparent and thus more difficult to push  via radiation
pressure. High flow-speeds are attained only in less ionized gas (that
might be farther downstream in the flow and partially shielded from
the intense ionizing radiation, e.g.,  Hamann \& Ferland, in prep). 

\begin{figure}
  \centering
  {\includegraphics[angle=0,width=80mm]{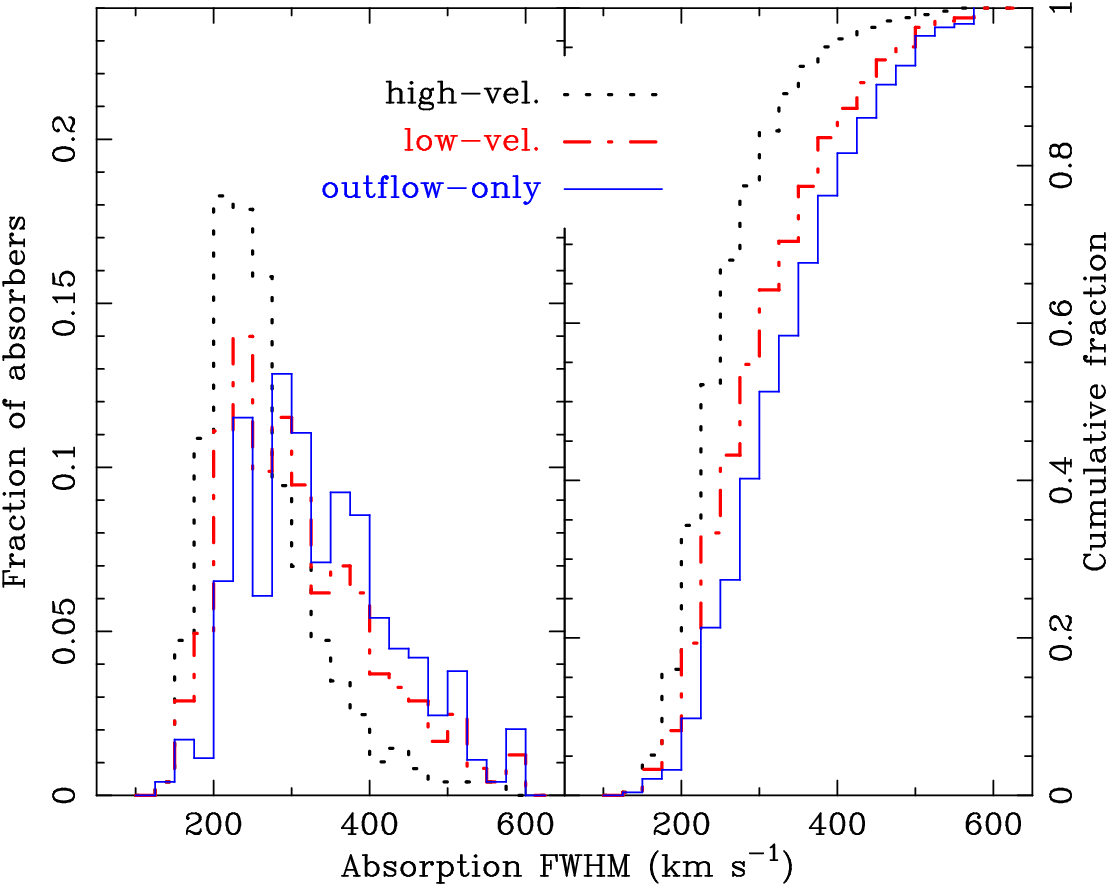}} 
\caption{The normalized (left) and cumulative (right) FWHM
distributions for high- and low-velocity absorber sub-samples.  The
low-velocity sample is clearly broader.  Note that the SDSS resolution
($\simeq 150$ \kms) sets the lower-limit seen in these data.  Also
shown (solid histograms) are the weighted difference distributions
(using the measured outflow fraction, see text), which represents the
distributions for outflowing-systems only.}
\label{Fig:fwhm}
\end{figure}

We can also infer the distribution of other outflow-absorber
properties by comparing the high-velocity (intervening)  absorbers to
those at lower  velocities where $f_{outflow} > 0$ but the incidence
of environmental systems is negligible.  To optimize such a
comparison, we require a range of velocity large enough to include a
significant number of outflowing absorbers, while restricting
ourselves to velocities where the outflow fraction is high.
Fig. \ref{Fig:fwhm} shows the normalized (left) and cumulative (right)
distributions  of FWHMs for absorbers with 1800 \kms\ $< v < $ 4400
\kms, over which the effective outflow fraction is
$\left<f_{outflow}\right> = 0.61 \pm 0.07$ and the environmental
contribution is small, and 30000 \kms\ $< v < $ 60000 \kms, where we
assume the absorbers to all be cosmologically intervening systems.
The low-velocity FWHM distribution is clearly broader than the
high-velocity distribution.  A K-S test gives a zero per cent
probability that the two sets of data are drawn from the same parent
distribution.  A broader distribution is expected if the low-velocity
sample contains outflowing absorbers.  We also show in
Fig. \ref{Fig:fwhm} the weighted (assuming $f_{outflow} = 0.61$),
renormalized difference between the two distributions, which should
represent the pure outflow distribution and has a mean
$\left<\mathrm{FWHM}\right> = 337$ \kms\ -- though we note that the
SDSS resolution imposes an artificial lower-limit to the measured FWHM
of $\approx 150$ \kms.  We also compared the distribution of
$\lambda1548:\lambda1551$ doublet-ratios ($DR$) for the two velocity
sub-samples.  The doublet ratio distributions are similar except that
the low-velocity sample exhibits relatively more completely saturated
absorbers: $26 \pm 3$ per cent with $DR > 0.9$  compared to $15 \pm 2$
per cent at high-velocity, which implies that a third ($33 \pm 6$ per
cent) of outflowing absorbers in this velocity range have $DR >
0.9$.  This result implies that outflow systems contain fewer weak
kinematically-distinct components that are unresolved in the SDSS
spectra than do intervening absorbers.  Although individual outflowing
narrow \civ\ absorbers appear similar to non-intrinsic systems in
low/medium-resolution data, they are on average broader and more often
completely saturated.

\begin{figure}
  \centering
  {\includegraphics[angle=0,width=80mm]{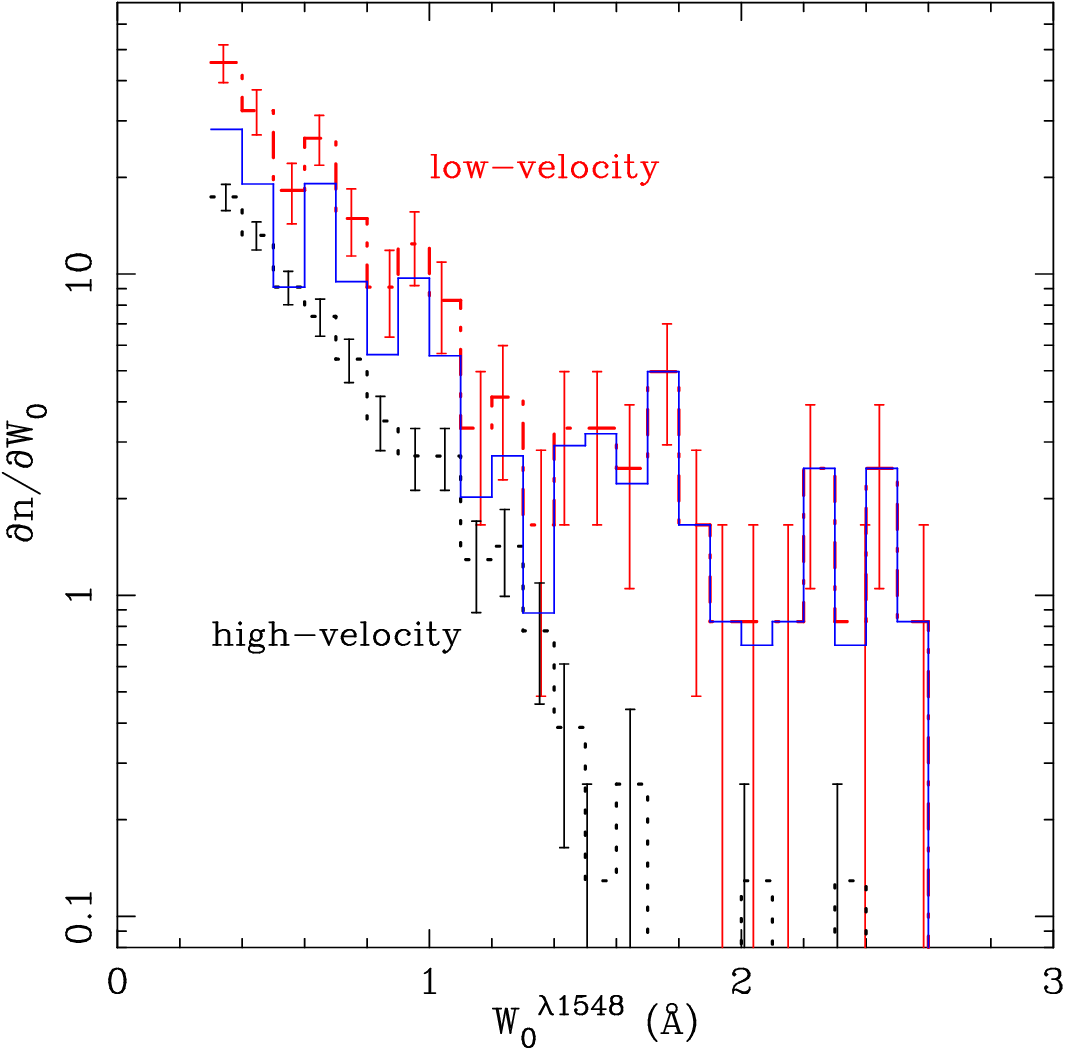}} 
  \caption{Incidence of absorbers as a function of \W, for high- and
    low-velocity samples.  Also shown (solid histogram)
 is the difference between the two distributions which represents
the distribution for outflow-systems only.}
  \label{Fig:dndw}
\end{figure}

Similarly, we can construct the \W\ distribution for the two velocity
ranges.  This is shown in Fig. \ref{Fig:dndw}. 
The distributions are normalized such that they represent the
differential number of absorbers per \AA\ per $3\times 10^5$ \kms\
averaged over the velocity range being considered.  The distribution
for outflow-only systems is therefore obtained simply from the
difference of these two distributions and is shown as the solid
histogram.  It is a much flatter distribution, with a very large
excess of strong systems compared to the intervening
absorbers. Notably, almost all of the low-velocity narrow absorbers
with \W $\ga 1.4$\AA\ can be attributed to outflowing systems.  These
differences are consistent with the broader FWHM distribution of
outflow-systems in Fig. \ref{Fig:fwhm}.

\subsection{How many quasars exhibit outflowing absorbers?}  
Although $\frac{\partial n}{\partial \beta}$ allows one to calculate
the average number of absorbers that will be found in a given range of
velocity space per quasar spectrum (i.e., Eqn. \ref{Eqn:dndv}), it
does not contain information on the {\it fraction of quasars} which
exhibit narrow \civ\ absorption.  We show in Table \ref{Tab:fracs} the
fraction of spectra that exhibit one or more and two or more narrow
\civ\ absorbers with \W\ $\ge 0.3$\AA\  for different velocity ranges,
as well as the average total number of systems per spectrum,
$\left<n\right>$. 
\begin{table}
  \caption{Fraction of quasars containing narrow \civ\ absorbers.}
\begin{center}
\label{Tab:fracs}
  \begin{tabular}{rcccccc}
\multicolumn{1}{c}{velocity range} & \multicolumn{3}{c}{all absorbers}
& \multicolumn{3}{c}{outflow-only systems}
\\ \multicolumn{1}{c}{(\kms)} & $1+$ & $2+$ & $\left<n\right>$ & $1+$
& $2+$ & $\left<n\right>$ \\
\hline
$ 1800 <v< 4400 $ & 0.13 & 0.03 &
0.17 & 0.08 & 0.03 & 0.11 \\ $ 0<v<3000 $      & 0.21 & 0.05 & 0.28 &
0.08 & 0.04 & 0.12 \\ $ 0<v<5000 $      & 0.25 & 0.08 & 0.36 & 0.09 &
0.05 & 0.16 \\ $ 0<v<12000 $     & 0.39 & 0.12 & 0.60 & 0.14 & 0.06 &
0.25 \\ $v<3000 $       & 0.27 & 0.08 & 0.37 &  &   &      \\  $v<5000
$       & 0.31 & 0.10 & 0.45 &  &   &      \\ 
\hline
\end{tabular}
\end{center}
The values represent the fraction of quasars that exhibit at least one
(1+) and at least two (2+) narrow \W\ $\ge 0.3$\AA\ \civ\ absorbers in
various velocity ranges, as well as the average number of absorbers
per spectrum.  Values for `all' absorbers are taken directly from the
data, while the fractions considering only `outflow' absorbers use our
fits to account for environmental and intervening systems, assuming
Poissonian statistics.
\end{table}

As we are primarily interested in the fraction of quasars exhibiting
absorbers that arise in outflows, we use our models for the incidence
of environmental and intervening systems as a function of velocity
(\S\ref{Sec:veldist} and Table \ref{Tab:fits}) to predict the fraction
of quasars having a given number of non-outflow absorbers for a given
velocity range, assuming Poissonian statistics.  While it is likely
that intervening absorbers exhibit some degree of clustering, any
such possible signal is at levels undetectable in our 
data (Nestor et al. 2007) and thus the Poissonian assumption is appropriate.
The predictions for
non-outflow absorbers can then be disentangled from the observed
fractions to calculate the fractions for outflowing systems only.
These fractions are also shown in Table \ref{Tab:fracs}.  In this
manner we estimate that approximately fourteen per cent of  the
quasars in our sample exhibit outflows that are detected in narrow
\civ\ absorption over the range $0 < v < 12000$ \kms, where the
intrinsic fraction is measured to be non-zero.

It is important to stress that the outflow-only fractions are minima,
since we may be counting some outflow-systems as non-outflow ones, and
they do not account for systems that we have eliminated due to
blending with broad (FWHM $>$ 600 \kms) absorption
(\S\ref{Sec:abssample}).  However, the total fractions (i.e., those
including the non-outflow absorbers) represent firm upper limits for
the outflow-only fractions.  Thus, the overall and calculated
outflow-only fractions represent windows in which the true
outflow-only fractions lie.  For the reasons described in
\S\ref{Sec:syst}, we expect that the true outflow-only fractions are
near the calculated outflow-only values presented in this table.  

\section{Summary and Implications}
\label{Sec:conc}
Our main new results are the following:

1) The distribution of narrow (FWHM $< 600$ \kms) \civ\ absorbers with
\W\ $\ge 0.3$\AA\ is well described by three  components: a Gaussian
`environmental' component  centred at $v=0$ \kms\ with a deconvolved
dispersion of  $\approx 450$ \kms, a flat cosmologically-intervening
component  at $v>0$ \kms\ with $\frac{\partial n}{\partial \beta}
\simeq 6.4$,  and an outflow component that extends from $v \simeq
750$ \kms\ to $v \simeq 12000$ \kms\ (Figs. \ref{Fig:dndv1} --
\ref{Fig:dndvi}). 

2) The outflow-systems account for a large fraction ($\ga$ 61 per
cent)  of all narrow-line absorbers in the velocity range  1800 \kms\
$\la v \la $ 4400 \kms.  The outflow fraction peaks near $v \simeq
2000$ \kms\ at $f_{outflow} \simeq 0.8$.  The true outflow fractions
might be even higher, and the tail of the distribution might reach
beyond 12,000 \kms , if some of the high-velocity narrow systems
(30000 \kms\ $\la v\la 60000$ \kms) also form  in outflows.  This
possibility is suggested by  results from previous studies (e.g.,
Richards et al.\ 2001; Misawa et al.\ 2007). 

3) The fraction of quasars with NAL outflows between 0 and 12,000
\kms\  is $\ga$14 per cent. The fraction of quasars with two or more
NAL outflow  systems at these velocities is $\ga$6 per cent. 

4) Among the narrow-line systems we consider (FWHM $< 600$ \kms),  the
outflowing absorbers (e.g., at 1800 \kms\ $\la v\la 4400$ \kms ) tend
to be broader than the intervening ones (at $v\ga 30000$ \kms).  The
outflow-systems also tend to be stronger and they are more  frequently
saturated, as measured by \W\ and the \civ\ doublet ratio.  

5) High resolution spectra are needed to examine the flow kinematics,
but the measured line widths $<$600 \kms\ imply that the projected
flow speeds  in these NAL outflows are characteristically $\sim$10
times larger than the line-of-sight velocity dispersions,  i.e.,
$v$/FWHM $\sim$ 10.

6) There appears to be a deficit of outflowing NALs at low velocities
($v\la 1500$ \kms\ compared to  1500\kms\ $\la v \la 8000$ \kms). 

These results provide important new constraints on models of quasar
outflows. For example, the fraction of quasars with  NAL outflows
($\ga$14 per cent) is at least as large as the fraction of  quasars
with classic (strong and broad) BALs ($\sim$10 per cent),  as measured
in similar optically-selected  quasar samples (Trump et al.\ 2006 and
refs. therein). Moreover, most of  the NAL systems in our study appear
in quasars without BALs, and we  specifically exclude NALs that are
blended with BALs. Therefore,  the NAL outflows we measure have no
obvious relationship to the BALs.  They might represent a wholly
different type of outflow phenomenon --  one that is perhaps less
massive but just as common as the BAL outflows.  Perhaps there is an
evolutionary relationship where massive BAL outflows  tend to
dissipate over time to become NAL outflows.  Alternatively, the
appearance of NAL outflow lines could be simply  a different
manifestation of a single quasar outflow phenomenon, where  our
viewing perspective through the flow  determines whether we see NALs
or BALs (e.g., Elvis et al.\  2000, Ganguly et al.\ 2001). Perhaps the
reality in quasars is a  mixture of these two scenarios (Hamann \&
Sabra 2004). 

The detection frequencies of the outflow lines provide   important
constraints on the flow structures.  In particular, the true fraction
of quasars  with BAL outflows, $f_B$, is related to the detection
frequency by  $f_B\ga 0.10$, while the true fraction with NAL outflows
is  $f_N\ga 0.14$. If there are no orientation biases that affect the
detection rates, then the covering factors of the outflows (as seen
from the central continuum source) are $\ga$0.14/$f_N$ for the NALs
and $\sim$0.10/$f_B$ for classic BALs. If all quasars have both NAL
and BAL outflows as part of a single outflow phenomenon, then the
covering factor of outflow gas is $\ga$0.24. These factors become even
larger if we include systems not counted here, e.g., the broader
mini-BALs  (Rodriguez Hidalgo et al., in prep) and the high-velocity
NALs (Misawa et al.  2007, Richards et al. 2001). Moreover, all of
these results based on  detection frequencies are lower limits if
quasars are characteristically  fainter when viewed along outflow
sightlines (Goodrich et al.\ 1997,  Schmidt \& Hines 1999), or if
quasars appear  obscured and are therefore missed by flux-limited
optical surveys when  viewed along any range of sightlines. 
On the other hand, if quasars are preferentially obscured 
along certain viewing angles by, e.g., a dusty torus, the above 
covering factors refer only to the unobscured solid-angle.

Further insight into the flow structures comes from the fact that
roughly half of the quasars with NAL outflows (6 of the total 14 per
cent) have two or more of these outflow features. Thus, the
line-of-sight  structure is often complex, with either multiple
streams or multiple  `clumps' contributing to the \civ\ absorption. In
addition, the NAL flow  speeds are in the low-to-medium range of BALs,
suggesting a similar  point of origin for these outflows, but the
large ratios of flow speed to line-of-sight velocity dispersion in
NALs, $v$/FWHM $\sim$ 10, contrast markedly with the BALs  where
typically $v$/FWHM $\sim$ 1. It is tempting to identify each  NAL
system with a clump of gas in the flow, but there is no compelling
model for these features. Clumps or filaments formed by instabilities
at  the ragged edge of a BAL flow should be short lived.  The measured
FWHMs of the NALs, when they are resolved in the SDSS  data at several
hundred \kms, suggest highly supersonic expansion. 

The possible deficit of low-velocity outflow NALs is another
interesting clue, but it also has no  straightforward
interpretation. This deficit might imply that  the inner flow is too
ionized for \civ , or that the  clumps or streams producing \civ\ NALs
do not form  in the inner flow where the speeds are low. There might
also  be orientation effects that preclude our detection of these
clumps or streams of \civ\ gas near the base of the flow. 

Although more work is certainly needed to resolve these questions,
NALs represent another important component  of quasar outflows.  Many
studies have focused on the BALs  because they are the most obvious
and most easily measured of the  UV outflow lines.  The overall
outflow phenomenon is clearly  much more diverse, however. Even the
NALs and BALs together do not tell the whole story  (Hamann \& Sabra
2004). These features are at opposite  extremes in terms of their
FWHMs, but in between is a  significant population of `mini-BALs'
which have intermediate  FWHMs and similar (or perhaps
characteristically higher)  outflow velocities compared to the NALs
and BALs (Rodriguez Hidalgo  et al., in prep). The challenge now is to
achieve a better physical  understanding of this full range of quasar
outflow features.

\label{lastpage}

\end{document}